\documentclass[final,5p,times,twocolumn,authoryear]{elsarticle}
\usepackage[utf8]{inputenc}
\usepackage[T1]{fontenc}
\usepackage{amsmath,amssymb}
\usepackage{graphicx}
\usepackage{booktabs}
\usepackage{hyperref}
\usepackage{bm}
\usepackage{array}
\usepackage{xcolor}

\journal{Journal of High Energy Astrophysics}

\begin{document}
\begin{frontmatter}

\title {Braking indices as probes of r-mode spin-down in young pulsars}

\author[UWO]{Shahram Abbassi\corref{cor1}}
\ead{sabbassi@uwo.ca}
\cortext[cor1]{Corresponding author}
\author[FUM]{Armin Memarian}
\author[UWO]{Evelyn Sophia Guest}
\author[UWO,KUC]{S.\,R.~Valluri}

\address[UWO]{Department of Physics and Astronomy,
              University of Western Ontario,
              London, ON N6A 3K7, Canada}
\address[FUM]{Department of Physics, Faculty of Sciences,
              Ferdowsi University of Mashhad,
              Mashhad 91775-1436, Iran}
\address[KUC]{Department of Mathematics, King's University College, University of Western Ontario, London, ON N6A 3K7, Canada}

\begin{abstract}
We present a timing-based framework for interpreting braking-index measurements in young pulsars within a four-channel spin-down model that includes particle-wind, magnetic-dipole, mass-quadrupole, and current-quadrupole torques. In this formulation, the observed braking index is a torque-weighted average of the channel exponents, allowing equation-of-state-independent constraints on the torque fraction of a possible r-mode-like current-quadrupole component from timing data alone. For a positive, slowly varying secular model, the physically allowed range is \(1\leq n_{\rm obs}\leq 7\). Within this domain, the lower and capped physical upper bounds on the \(k=7\) torque fraction are \(f_{7,\min}=\max[0,(n_{\rm obs}-5)/2]\) and \(f_{7,\max}^{\rm phys}=\min\{1,\max[0,(n_{\rm obs}-1)/6]\}\), respectively. These bounds are conditional on the non-negativity and slow secular evolution of the effective torque coefficients; if these assumptions are relaxed, a large braking index need not uniquely imply a current-quadrupole torque. Applying this bracket analysis to young pulsars with measured braking indices, we find that most sources do not require a gravitational-wave torque and can be explained by wind-plus-dipole electromagnetic spin-down. Sources with \(3<n_{\rm obs}<5\) require some higher-order contribution beyond the wind and dipole terms, but timing alone cannot uniquely identify this contribution as an r-mode torque. PSR~J0537$-$6910 remains the most suggestive case: its large inter-glitch braking index approaches the \(n\simeq 7\) limit, where the timing bounds are consistent, within the restricted positive secular model, with a strong current-quadrupole-like contribution. This interpretation is not unique, however, because vortex-creep and superfluid-recovery effects can generate similar effective inter-glitch braking behaviour without invoking gravitational-wave emission. We also translate the timing constraints into stellar-model-dependent r-mode amplitude bounds, braking-index-corrected characteristic ages, and gravitational-wave ranking metrics for continuous-wave searches with current and future detectors, including the reduction in sensitivity expected for glitch-limited semi-coherent searches.
\end{abstract}


\end{frontmatter}

\section{Introduction}
\label{sec:intro}

The long-term spin evolution of neutron stars offers one of the clearest observational windows into the interplay between dense matter, magnetic structure, superfluid dynamics, and gravitational-wave emission. Young pulsars, in particular, evolve on timescales short enough that their measured rotational parameters preserve valuable information about the torques that have acted over their secular evolution. That information is encoded in the spin frequency $\nu$, its first derivative $\dot{\nu}$, and the braking index $n \equiv \nu\ddot{\nu}/\dot{\nu}^{2}$. Because the braking index depends on both the magnitude and frequency scaling of the secular torque, it has long been regarded as a key diagnostic of neutron-star spin-down physics.

In the idealized picture of pure magnetic-dipole braking, $\dot{\nu} \propto -\nu^{3}$ and therefore $n=3$ \citep{LorimerKramer2004}. Observations have shown, however, that this simple picture is incomplete. Measured braking indices in young pulsars span a broad range, from the sub-unity value $n \simeq 0.9$ for PSR~J1734$-$3333 \citep{Espinoza2011} to values above the dipole expectation, such as $n \simeq 3.15$ for PSR~J1640$-$4631 \citep{Archibald2016}. Several systems also exhibit secular evolution, timing-noise contamination, glitch-related torque offsets, or magnetospheric state changes \citep{Livingstone2007,Parthasarathy2019,Espinoza2017}. Recent studies have further emphasized that magnetic-field evolution, inclination-angle evolution, particle winds, and gravitational-wave torques can all modify the effective braking index and therefore complicate any single-torque interpretation \citep{Gao2016_MagnetarBraking,Gao2017_J1640,Li2025_J1846Inclination,Li2026_SwiftJ1834}. These measurements strongly suggest that the spin-down of young pulsars is, in general, multi-channel rather than governed by a single stationary mechanism.

This point is especially important in the context of gravitational-wave astrophysics. For isolated pulsars, two gravitational-wave channels are of particular interest. A static mass quadrupole produces $\dot{\nu} \propto -\nu^{5}$ and hence $n=5$ \citep{Owen2005,Ushomirsky2000}, while a current quadrupole associated with unstable r-mode oscillations drives a steeper dependence, $\dot{\nu} \propto -\nu^{7}$, corresponding asymptotically to $n \to 7$ \citep{Andersson1998_rmode,FriedmanSchutz1978,LindblomOwenMorsink1998,Owen1998}. If present at an astrophysically relevant level, these channels would not only modify the timing evolution of young pulsars, but could also make them promising continuous-wave targets for current and future detectors. At the same time, a large measured or effective braking index is not, by itself, a unique signature of gravitational-wave emission, because time-dependent electromagnetic torques, magnetic-field decay, magnetospheric transitions, or internal angular-momentum exchange can also alter the measured value of $n$.

Among these possibilities, r-modes are particularly compelling. They arise from the Chandrasekhar--Friedman--Schutz radiation-reaction instability and can, in principle, convert rotational energy efficiently into gravitational radiation \citep{Andersson1998_rmode,FriedmanSchutz1978,AnderssonKokkotas2001}. At the same time, their astrophysical relevance remains highly model dependent. The instability window is shaped by competing dissipation mechanisms, including shear viscosity, bulk viscosity, crust--core boundary-layer damping, and superfluid mutual friction \citep{HaskellAndersson2015,GusakovChugunov2019,Kraav2024}. Magnetic effects may further alter the boundaries \citep{Abbassi2012}, and theoretical estimates of the saturation amplitude still span several orders of magnitude, roughly $10^{-8} \lesssim \alpha_{\rm sat} \lesssim 10^{-4}$ \citep{Arras2003_saturation,Bondarescu2007,MahmoodifarStrohmayer2013,GusakovChugunov2019,AlfordSchwenzer2014a}. Observationally, the current LIGO~O3 upper limit for PSR~J0537$-$6910 already probes part of this parameter space \citep{Abbott2021_J0537_rmode}. The question is therefore not merely whether r-modes are theoretically allowed, but whether timing data can help identify where an r-mode interpretation is plausible, where it is disfavoured, and where it remains degenerate with non-radiative internal dynamics.

A major practical difficulty is that braking-index analyses are often framed in terms of a single dominant torque, even though the data themselves may reflect a superposition of physically distinct channels. In Paper~I \citep{LiAbbassi2025}, we introduced a four-term spin-down model,
\begin{displaymath}
\dot{\nu} = -s\nu - r\nu^{3} - g\nu^{5} - \ell\nu^{7},
\end{displaymath}
and showed that it can reproduce the secular spin evolution of the Crab pulsar within a unified phenomenological framework. The present paper takes a complementary step. Rather than fitting the four coefficients source by source, we ask what can be inferred directly from a measured braking index under the assumption that the spin-down is described by this four-channel model with non-negative, slowly varying effective coefficients. This assumption is central to the interpretation: if the effective coefficients vary rapidly, or if an additional spin-up or recovery term contributes over the measured interval, the braking index no longer has to behave as a simple convex average of the four channel exponents.

Within this formalism, the braking index reduces to a torque-weighted average of the channel exponents. This identity leads to timing-based bounds on the fractional contribution of the $k=7$ current-quadrupole channel. It also identifies when the observed braking index can be accommodated within the electromagnetic wind-plus-dipole sector and when an additional higher-order torque is required. This distinction is central to the interpretation. Timing can reveal the need for a steeper torque component, but it does not by itself uniquely identify that component as an r-mode contribution in every source. In particular, sources with $3<n<5$ require a higher-order contribution beyond the electromagnetic sector within the present model, but that contribution may in principle be supplied by a mass-quadrupole channel rather than by an r-mode. Only for $n>5$ does the positive four-channel model imply a strictly nonzero lower bound on the $k=7$ fraction. This statement should be read as a conditional result of the positive, slowly varying secular model, not as a model-independent proof of r-mode emission.

The motivation for this work is therefore twofold. First, we seek a transparent and observationally economical way to connect braking-index measurements to multi-channel spin-down physics without requiring full source-specific torque fitting. Second, we aim to identify those pulsars for which an r-mode interpretation is genuinely worth pursuing in concert with gravitational-wave searches. In that sense, the novelty of the present paper lies not in proposing a new microscopic model of r-mode saturation, but in constructing a timing-based diagnostic framework that bridges pulsar phenomenology and continuous-wave astronomy in a way that is simple, testable, and readily applicable to existing data. A useful outcome of such a framework is not only to identify possible r-mode candidates, but also to state clearly what observations would weaken or falsify the r-mode interpretation.

Our starting literature census contains 14 young pulsars with published braking-index measurements. The formal torque-fraction bracket analysis is restricted to the 13 sources with $1\leq n_{\rm obs}\leq 7$, which is the domain implied by non-negative torque fractions in the four-channel secular model. PSR~J1734$-$3333, with $n_{\rm obs}\simeq 0.9$, is therefore treated as an excluded boundary case. Its sub-unity braking index is nevertheless physically instructive, since it points to secular magnetic-field evolution, magnetospheric changes, or a breakdown of the slowly varying positive-torque assumption rather than to a steady combination of the four dissipative channels considered here.

Applying the framework to the 13 sources in the formal domain, we find that most require no gravitational-wave torque and can be accommodated by wind-plus-dipole electromagnetic spin-down. A smaller subset requires higher-order torque contributions beyond the electromagnetic sector. PSR~J0537$-$6910 emerges as the most interesting case: its near-asymptotic inter-glitch braking index approaches $n\simeq 7$, placing it in the regime most strongly compatible, within the restricted positive secular model, with a strong current-quadrupole contribution. This makes J0537 the most promising source in our sample for a timing-based r-mode interpretation, although viable superfluid-dynamical alternatives, including vortex-creep recovery, can produce similar effective inter-glitch behaviour without requiring gravitational-wave emission. For this reason, J0537 is treated here as a test case for distinguishing an r-mode torque from internal recovery physics, rather than as a unique detection of r-mode spin-down from timing data alone. We further translate the timing constraints into bounds on r-mode amplitude, braking-index-corrected characteristic ages, and gravitational-wave ranking measures relevant to advanced and third-generation interferometers. Because J0537 glitches frequently, we distinguish the idealized year-long coherent ranking from an optimistic 30-day coherence-limited estimate, while emphasizing that a realistic semi-coherent search will generally be less sensitive.

The constraints developed here are algebraically independent of the neutron-star equation of state at the level of torque fractions, but their translation into mode amplitudes, strain estimates, and detectability rankings necessarily inherits stellar-model systematics. The resulting framework should therefore be viewed as a phenomenological diagnostic: observationally grounded and directly testable, but conditional on the assumed positive four-channel secular model. Its main purpose is to separate what timing data alone can establish robustly from what requires additional physical modeling, thermal information, or gravitational-wave observations. This separation is especially important when comparing the r-mode interpretation with alternatives such as magnetic-field evolution, magnetospheric state changes, glitch recovery, and vortex-creep relaxation.

The paper is organised as follows. Section~\ref{sec:formalism} introduces the torque-fraction formalism. Section~\ref{sec:bounds} derives the timing-based constraints on the $k=7$ torque fraction. Section~\ref{sec:rmode_physics} connects these constraints to r-mode amplitudes, strain estimates, braking-index-corrected ages, the young-pulsar census, the J0537 inter-glitch analysis, and gravitational-wave ranking metrics. Section~\ref{sec:discussion} discusses physical interpretation, limitations, and implications for gravitational-wave searches. Section~\ref{sec:conclusions} summarises the main results.

\section{Spin-down formalism}
\label{sec:formalism}

\subsection{Four-channel model}
\label{sec:model}

The secular spin evolution of an isolated neutron star is generally the result of several torque channels acting simultaneously. Each channel reflects a distinct physical mechanism and carries a characteristic dependence on the spin frequency. A convenient phenomenological starting point is the requirement of rotation-reversal antisymmetry: under $\nu\rightarrow-\nu$, the torque must also reverse sign \citep{BlandfordRomani1988}. This symmetry motivates a spin-down law written in odd powers of $\nu$,
\begin{equation}
\dot{\nu}
=
-\sum_{k=1,3,5,7,\ldots} c_k \nu^k,
\label{eq:general}
\end{equation}
where the coefficients $c_k$ encode the effective strengths of the corresponding torque channels.

In this work, we do not regard Eq.~(\ref{eq:general}) as a strictly
convergent expansion controlled only by powers of $\nu R/c$, since
the dimensional coefficients $c_k$ represent physically distinct
mechanisms and can differ substantially from one channel to another.
Instead, we adopt a minimal phenomenological truncation that retains
the four torque scalings most commonly associated with young-pulsar
spin-down and gravitational-wave emission.

The truncation at $k=7$ is motivated by the physical content
of the retained terms. The $k=1$, $3$, $5$, and $7$ contributions
represent, respectively, particle-wind braking, magnetic-dipole
braking, mass-quadrupole gravitational-wave emission, and
current-quadrupole emission associated with r-mode-like dynamics.
These channels comprise the standard secular mechanisms most
commonly invoked in the spin evolution of isolated young pulsars
and include both the dominant electromagnetic contributions and
the leading gravitational-wave torque scalings relevant to the
present problem.

We do not claim that all higher-order effective torques
vanish. Rather, no comparably established persistent spin-down
mechanism for an isolated young neutron star is conventionally
associated with a leading $k\geq9$ dependence. The four-channel
model should therefore be understood as the minimal physically
motivated framework needed to bracket the possible $k=7$
contribution. If an additional higher-order secular torque were
dynamically important, the derived bounds would need to be
generalized accordingly.

The resulting model is
\begin{equation}
\dot{\nu}
=
-s\nu-r\nu^3-g\nu^5-\ell\nu^7,
\label{eq:nudot}
\end{equation}

where $s$, $r$, $g$, and $\ell$ are taken to be non-negative and slowly varying on secular timescales. The four retained terms are associated, respectively, with particle-wind losses, magnetic-dipole braking, mass-quadrupole gravitational-wave emission, and current-quadrupole emission associated with r-mode-like dynamics. Their corresponding frequency exponents are $k=1$, $3$, $5$, and $7$, giving isolated single-channel braking indices $n=1$, $3$, $5$, and $7$.

The non-negativity assumption is a physical restriction on the class of models considered here: each retained term is treated as a dissipative spin-down channel. It excludes, by construction, transient spin-up contributions, short-lived recovery torques, and sign-changing effective terms that may arise during glitch relaxation or magnetospheric switching. If such terms are important over the interval used to measure the braking index, the observed value of $n$ should not be interpreted as a simple torque-weighted average of the four channel exponents.

The slowly varying assumption can be stated more quantitatively by requiring the fractional rate of change of each effective coefficient to be small compared with the fractional spin-down rate over the timing baseline used to infer $n$, namely
\begin{equation}
\left|\frac{d\ln c_k}{dt}\right|
\ll
\left|\frac{\dot{\nu}}{\nu}\right|.
\label{eq:slow_condition}
\end{equation}
This condition is intended as a sufficient secular criterion rather than as a claim that every young pulsar satisfies it at all times. When Eq.~(\ref{eq:slow_condition}) fails, the measured braking index contains additional terms proportional to $dc_k/dt$, and the clean algebraic bounds derived below no longer apply without further modelling.

The assumption of slowly varying coefficients is essential for the interpretation. Over the interval on which a braking index is measured, the parameters in Eq.~(\ref{eq:nudot}) should be understood as effective secular coefficients. When this assumption breaks down, for example during glitch recovery, magnetospheric state changes, secular magnetic-field evolution, or other strongly time-dependent episodes, the model no longer describes the instantaneous microphysics. In such cases, it should be read only as an effective time-averaged torque decomposition, or the source may fall outside its domain of applicability altogether. This caveat is particularly important for sources such as PSR~J0537$-$6910, where the inter-glitch braking index may approach the $k=7$ scaling, but where vortex-creep recovery and other internal relaxation processes can also generate a large effective braking index without requiring a gravitational-wave torque.

For PSR~J0537$-$6910, the magnitude of a possible violation can be estimated directly. Using its characteristic age, $\tau_c\simeq4.9\,\mathrm{kyr}$, the secular fractional spin-down rate is
\begin{equation}
\left|\frac{\dot{\nu}}{\nu}\right|
=
\frac{1}{2\tau_c}
\simeq
3.2\times10^{-12}\,\mathrm{s}^{-1}.
\label{eq:J0537_secular_rate}
\end{equation}
If an effective torque coefficient changes by a fractional amount $\epsilon_{\rm rec}$ over a characteristic post-glitch recovery time $t_{\rm rec}\sim10\,\mathrm{d}$, then
\begin{equation}
\left|\frac{d\ln c_k}{dt}\right|
\sim
\frac{\epsilon_{\rm rec}}{t_{\rm rec}}
\simeq
1.2\times10^{-6}\epsilon_{\rm rec}\,\mathrm{s}^{-1}.
\label{eq:J0537_recovery_rate}
\end{equation}
The ratio of the recovery rate to the secular spin-down rate is therefore
\begin{equation}
\frac{\left|d\ln c_k/dt\right|}
     {\left|\dot{\nu}/\nu\right|}
\sim
3.7\times10^{5}\,\epsilon_{\rm rec}.
\label{eq:J0537_rate_ratio}
\end{equation}
Thus, an order-unity coefficient variation would violate the slow-variation condition by about five orders of magnitude, while even a one-percent variation over ten days would exceed the secular rate by approximately $3.7\times10^{3}$. The condition in Eq.~(\ref{eq:slow_condition}) is therefore strongly violated during rapid post-glitch recovery. The four-channel interpretation is most defensible only during the late inter-glitch interval, after the dominant recovery terms have decayed and the timing evolution approaches a quasi-stationary regime. Even in that interval, the assumption should be regarded as an approximation rather than as a demonstrated property of the source.
\subsection{Torque fractions and braking-index identity}
\label{sec:identity}

To express the model in dimensionless form, we define the total spin-down rate coefficient
\begin{equation}
T_{\rm tot}(\nu) \equiv -\frac{\dot{\nu}}{\nu}
= s + r\nu^2 + g\nu^4 + \ell\nu^6 ,
\label{eq:Ttot}
\end{equation}
together with the four torque fractions
\begin{equation}
f_{1} \equiv \frac{s}{T_{\rm tot}},\quad
f_{3} \equiv \frac{r\nu^{2}}{T_{\rm tot}},\quad
f_{5} \equiv \frac{g\nu^{4}}{T_{\rm tot}},\quad
f_{7} \equiv \frac{\ell\nu^{6}}{T_{\rm tot}} .
\label{eq:fracs}
\end{equation}
For dissipative torque contributions these fractions satisfy
\begin{displaymath}
f_1+f_3+f_5+f_7=1,\qquad f_k\geq0 .
\end{displaymath}
They therefore provide a convenient measure of the fractional contribution of each channel to the total secular spin-down at a given frequency.

Using Eq.~(\ref{eq:nudot}), and assuming that the coefficients are constant or slowly varying over the timing interval used to measure the braking index, the braking index
$n=\nu\ddot{\nu}/\dot{\nu}^{2}$ becomes
\begin{equation}
n(\nu) =
\frac{s + 3r\nu^2 + 5g\nu^4 + 7\ell\nu^6}
{s + r\nu^2 + g\nu^4 + \ell\nu^6}.
\label{eq:n_full}
\end{equation}
After substituting the definitions in Eq.~(\ref{eq:fracs}), this reduces to
\begin{equation}
n = f_{1} + 3f_{3} + 5f_{5} + 7f_{7}.
\label{eq:n_identity}
\end{equation}
Equation~(\ref{eq:n_identity}) shows that, within the positive secular four-channel model, the observed braking index is the torque-weighted mean of the channel exponents ${1,3,5,7}$. As an immediate consequence of the positivity of the fractions, the model implies
\begin{equation}
1 \leq n \leq 7 .
\label{eq:n_domain}
\end{equation}

It is useful to make explicit where the slow-variation assumption enters. If the coefficients $c_k$ in Eq.~(\ref{eq:general}) vary appreciably with time, differentiating the general spin-down law gives an additional contribution to the measured braking index,
\begin{equation}
n =
\sum_k k f_k
- \frac{1}{T_{\rm tot}}
\sum_k f_k \frac{d\ln c_k}{dt}.
\label{eq:n_time_dependent}
\end{equation}
The simple weighted-average identity in Eq.~(\ref{eq:n_identity}) is therefore recovered only when the second term in Eq.~(\ref{eq:n_time_dependent}) is negligible. This is precisely the secular condition stated in Eq.~(\ref{eq:slow_condition}). If magnetic-field evolution, magnetospheric switching, glitch recovery, or any other rapidly varying process contributes significantly over the timing baseline, the measured braking index need not lie in the interval $1\leq n\leq7$ and should not be interpreted as a direct torque-fraction constraint.

The interval in Eq.~(\ref{eq:n_domain}) is therefore not a generic prediction for all pulsars. It is the domain of validity of the positive, slowly varying four-channel secular model adopted here. Braking indices outside this range should not be forced into the torque-fraction interpretation. Values $n<1$ may point to secular magnetic-field evolution, magnetospheric changes, or an effective breakdown of the positive-coefficient assumption. Values $n>7$, which can occur as transient inter-glitch or post-glitch effective indices, are more naturally interpreted as evidence for rapidly varying internal or magnetospheric dynamics rather than as steady torque fractions within Eq.~(\ref{eq:nudot}).

Equation~(\ref{eq:n_identity}) is the central algebraic result of the present formalism. It makes clear that a measured braking index should not, in general, be interpreted as evidence for a single dominant mechanism. Rather, $n$ represents the collective imprint of all active channels. At the same time, Eq.~(\ref{eq:n_identity}) also shows the main limitation of a single braking-index measurement: it constrains only one linear combination of the four fractions. Timing data alone therefore cannot uniquely isolate the r-mode contribution without additional inequalities or modelling assumptions. What can be inferred robustly from Eq.~(\ref{eq:n_identity}) is developed in Section~\ref{sec:bounds}.

\begin{figure}[!ht]
\centering
\includegraphics[width=\columnwidth]{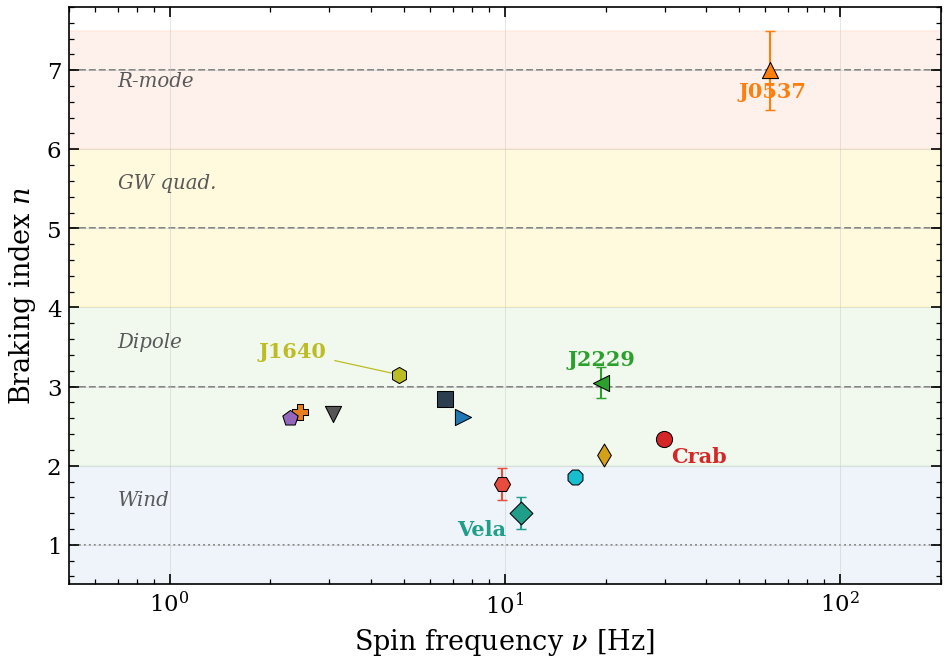}
\caption{Census pulsars plotted at their observed
$(\nu,n_{\rm obs})$. Horizontal dashed lines mark the isolated
single-channel values $n=1,3,5,$ and $7$. The shaded bands
indicate the braking-index regimes implied by
Eq.~(\ref{eq:n_identity}). They should be interpreted as
diagnostic intervals within the positive four-channel
secular model, not as unique identifications of the
underlying torque mechanism. In practice, the observed
braking index reflects a weighted superposition of all
active channels, and error bars on $n_{\rm obs}$ reflect
published uncertainties where available.}
\label{fig:n_vs_nu}
\end{figure}

Figure~\ref{fig:n_vs_nu} presents the census pulsars at their observed $(\nu,n_{\rm obs})$ within the braking-index regimes implied by Eq.~(\ref{eq:n_identity}).

\section{Timing constraints}
\label{sec:bounds}

The braking-index identity in Eq.~(\ref{eq:n_identity}), together with the positivity conditions $f_k\geq0$ and $f_1+f_3+f_5+f_7=1$, allows a limited but useful set of source-independent inferences about the $k=7$ torque fraction. When this channel is identified with an r-mode current quadrupole, these inferences become timing-based constraints on the r-mode contribution to the secular spin-down. The strongest result is an upper bound that follows directly from timing data alone. A lower bound can also be obtained, but its interpretation requires more care: in the full four-channel model, a measured braking index above $3$ implies the presence of a higher-order torque beyond the wind-plus-dipole electromagnetic sector, but it does not by itself require a nonzero r-mode contribution. All bounds in this section are therefore conditional on the positive, slowly varying secular assumptions stated in Section~\ref{sec:model}. If negative, sign-changing, or rapidly time-dependent effective torques are allowed, the algebraic interpretation of $n$ as a convex average of the four exponents is no longer valid.

\subsection{Upper bound}
\label{sec:upper}

At fixed observed braking index $n_{\rm obs}$, the largest possible value of $f_7$ is obtained by assigning the remaining torque fraction to the lowest-exponent channel, namely $f_1$, and setting $f_3=f_5=0$. Equation~(\ref{eq:n_identity}) then gives
\begin{equation}
f_{7,\max} = \frac{n_{\rm obs}-1}{6}.
\label{eq:f7max}
\end{equation}
This formal bound is saturated when the spin-down is shared exclusively between the wind and $k=7$ channels. Within the physical domain of the positive four-channel model, however, the torque fraction cannot exceed unity. We therefore define the capped physical upper bound
\begin{equation}
f_{7,\max}^{\rm phys}=
\min\left[
1,
\max\left(0,\frac{n_{\rm obs}-1}{6}\right)
\right].
\label{eq:f7max_phys}
\end{equation}

For sources with $1\leq n_{\rm obs}\leq7$, Eq.~(\ref{eq:f7max_phys}) coincides with Eq.~(\ref{eq:f7max}). For effective braking indices outside this interval, the capped expression prevents an unphysical interpretation in terms of $f_7>1$ or $f_7<0$ and signals that the steady four-channel assumptions should be re-examined.

For the Crab pulsar ($n_{\rm obs}=2.342\pm0.001$; \citealt{Lyne2015}), Eq.~(\ref{eq:f7max}) gives $f_{7,\max} \simeq 0.224$. For PSR~J0537$-$6910, whose near-asymptotic inter-glitch braking index approaches $n_{\rm obs}\to7$, the physical upper bound tends to $f_{7,\max}^{\rm phys}\to1$.

The propagated uncertainty in the formal bound is simply
\begin{equation}
\sigma_{f_{7,\max}} = \frac{\sigma_n}{6},
\label{eq:f7max_error}
\end{equation}
which is negligible for well-measured sources ($\sigma_n\lesssim0.01$) and remains only at the few-percent level even for timing-noise-dominated systems ($\sigma_n\sim0.1$--$0.5$). This statistical uncertainty should not be confused with the larger systematic uncertainty involved in interpreting the $k=7$ channel specifically as an r-mode torque. In particular, Eq.~(\ref{eq:f7max}) is a bound on the allowed current-quadrupole-like torque fraction inside the adopted secular model; it is not a direct measurement of an r-mode amplitude or gravitational-wave luminosity.

\subsection{Lower bound}
\label{sec:lower}

A robust lower bound on $f_7$ can also be derived directly from Eq.~(\ref{eq:n_identity}), but it is weaker than one might naively infer from the condition $n_{\rm obs}>3$. Rearranging Eq.~(\ref{eq:n_identity}) by substituting
$f_5=1-f_1-f_3-f_7$ gives
\begin{equation}
n-5 = -4f_1 - 2f_3 + 2f_7 .
\label{eq:n_minus_five}
\end{equation}
Since $f_1\geq0$ and $f_3\geq0$, it follows that
\begin{equation}
n-5 \leq 2f_7 ,
\label{eq:f7_ineq}
\end{equation}
and therefore
\begin{equation}
f_{7,\min}=
\max\left[
0,\frac{n_{\rm obs}-5}{2}
\right].
\label{eq:f7min}
\end{equation}

Combining Eqs.~(\ref{eq:f7max_phys}) and (\ref{eq:f7min})
yields the physical timing-based bracket
\begin{equation}
f_{7,\min} \leq f_7 \leq f_{7,\max}^{\rm phys}.
\label{eq:bracket}
\end{equation}

This bracket clarifies the physical meaning of different braking-index regimes. For $n_{\rm obs}<3$, the observed spin-down can be accommodated within the wind-plus-dipole electromagnetic sector, and timing places only an upper bound on $f_7$. For $3<n_{\rm obs}<5$, an additional higher-order torque is required beyond the electromagnetic sector; however, that contribution may be supplied entirely by the $k=5$ mass-quadrupole channel, so timing alone does not require $f_7>0$ in this interval. Only when $n_{\rm obs}>5$ does the positive four-channel model imply a strictly nonzero lower bound on the $k=7$ fraction. In the limiting case $n_{\rm obs}=7$, the steady model approaches $f_7\to1$.

The last statement is deliberately model-conditional. It does not mean that every pulsar with an effective braking index larger than $5$ must be losing rotational energy through r-mode gravitational radiation. Rather, it means that, within the restricted class of non-negative and slowly varying dissipative torques considered here, an exponent steeper than the mass-quadrupole value cannot be reproduced without a nonzero $k=7$ contribution. If magnetic-field decay, magnetospheric state changes, glitch recovery, vortex-creep relaxation, or a transient spin-up torque contributes to the measured $n_{\rm obs}$, the inference of a positive $f_7$ no longer follows from timing alone.

This distinction is important for the interpretation of the census sources. Pulsars such as J1640$-$4631 and J2229+6114, with braking indices slightly above $3$, require a higher-order contribution beyond the electromagnetic sector, but they do not demand a nonzero r-mode torque fraction from timing data alone. By contrast, a source with $n_{\rm obs}>5$ would lie in a regime where a nonzero $f_7$ is unavoidable within the present formalism. Observationally, such a case should be treated as a candidate for follow-up rather than as a unique r-mode identification; additional information from glitch recovery, timing-noise statistics, thermal state, and continuous-wave searches is needed to distinguish the r-mode interpretation from non-radiative alternatives.

\begin{figure*}[!ht]
\centering
\includegraphics[width=1.85\columnwidth]{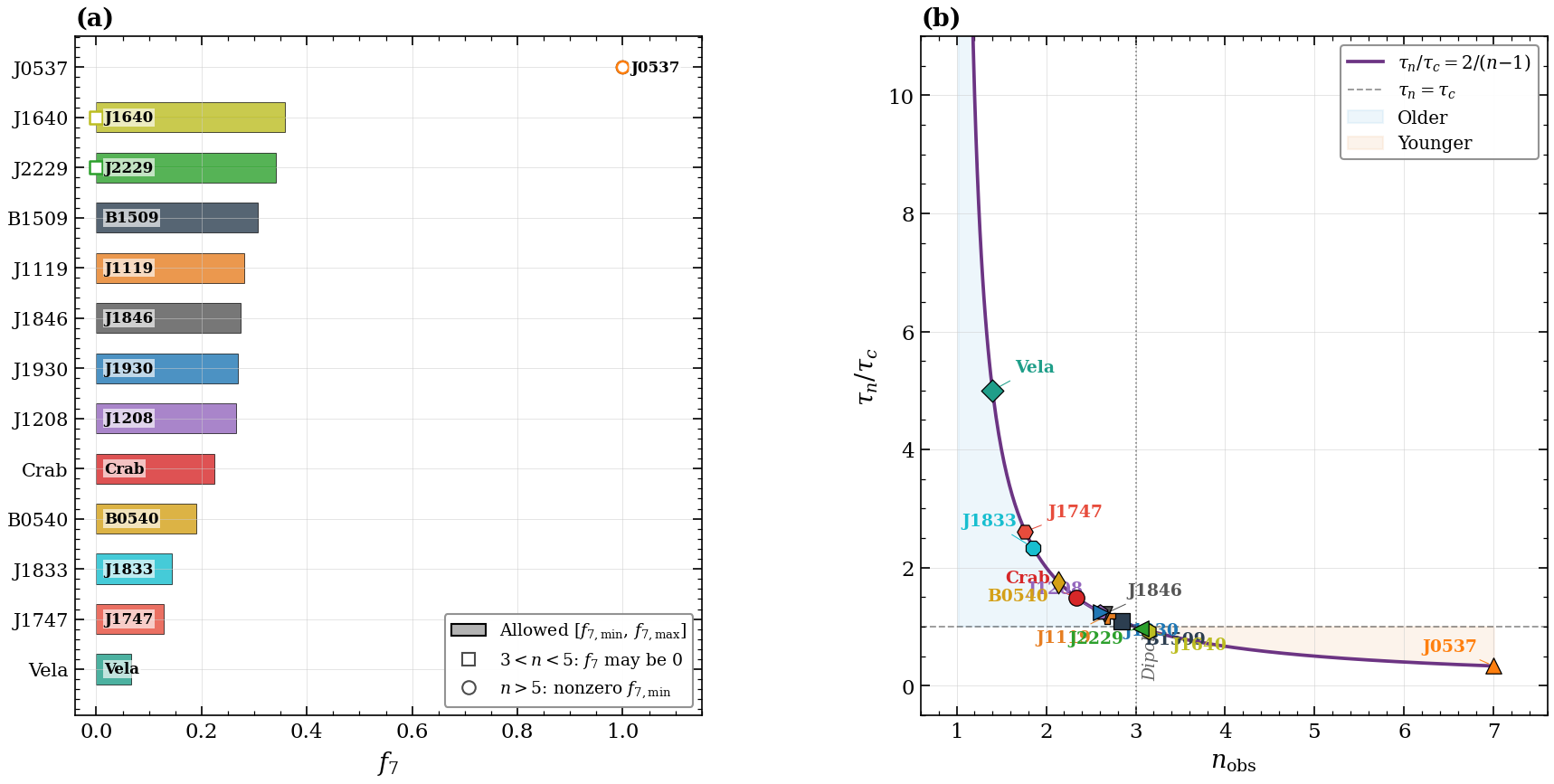}
\caption{(a) Timing-based $f_7$ brackets for the census
pulsars, ordered by braking index. Horizontal bars span the
physical interval $[f_{7,\min},f_{7,\max}^{\rm phys}]$. For
sources with $3<n_{\rm obs}<5$, a square marker at $f_7=0$
indicates that a higher-order torque is required, but the
$k=7$ fraction may still vanish. For $n_{\rm obs}>5$ (J0537
only), a circular marker shows the nonzero robust lower
bound. (b) Braking-index-corrected age ratio
$\tau_n/\tau_c=2/(n_{\rm obs}-1)$ for each source. Sources above
the horizontal line are older than the standard dipole
estimate, while sources below it are younger.}
\label{fig:two_sided}
\end{figure*}

\subsection{The $n=5$ threshold}
\label{sec:gw_threshold}

Additional intuition can be gained by considering the restricted gravitational-wave-dominated limit, in which the electromagnetic channels are neglected and $f_1=f_3=0$ with $f_5+f_7=1$. In that case, Eq.~(\ref{eq:n_identity}) reduces to
\begin{equation}
f_7 = \frac{n-5}{2}, \qquad
f_5 = \frac{7-n}{2}.
\label{eq:gw_channel}
\end{equation}
Within this restricted limit, the value $n=5$ marks the transition between mass-quadrupole dominance ($n<5$) and current-quadrupole dominance ($n>5$). Among the sources in the formal census, only PSR~J0537$-$6910 lies above this threshold.

This threshold is useful because it identifies the point at which the steeper gravitational-wave-like channel overtakes the mass-quadrupole contribution in the restricted $f_5+f_7=1$ subspace. It should not, however, be over-interpreted: the inference applies only to this gravitational-wave-dominated limit and not to the full four-channel problem without additional assumptions. In particular, the threshold does not by itself distinguish a genuine r-mode torque from other processes that can generate a large effective braking index, such as vortex-creep relaxation, glitch recovery, secular magnetic-field evolution, or magnetospheric state changes.

\begin{figure}[!ht]
\centering
\includegraphics[width=\columnwidth]{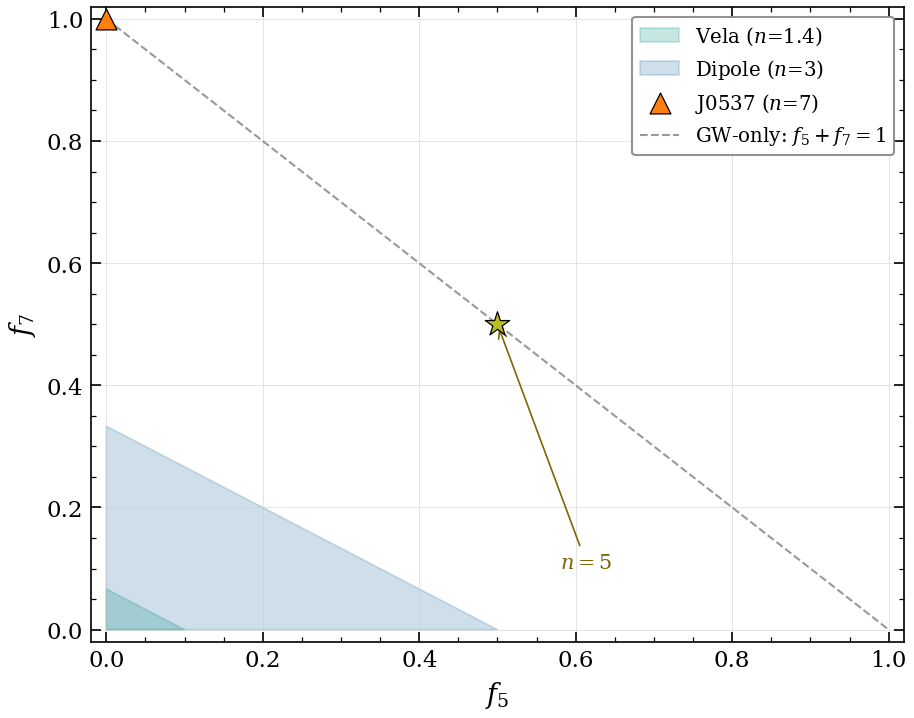}
\caption{Allowed regions in the $(f_5,f_7)$ plane for
three representative braking indices: Vela, the dipole
value, and J0537. The regions are subject to $f_k\geq0$.
The dashed diagonal marks the restricted
gravitational-wave-dominated limit $f_5+f_7=1$. The star at
$n=5$ separates mass-quadrupole-dominated and
current-quadrupole-dominated behaviour within that
restricted limit. As $n$ increases from 1.4 to 7, the
allowed region shifts systematically toward the $f_7$ axis.}
\label{fig:f5f7}
\end{figure}

Figure~\ref{fig:f5f7} illustrates the allowed $(f_5,f_7)$ regions for representative braking indices.

\section{Amplitude, strain, and ages}
\label{sec:rmode_physics}

\subsection{Amplitude bounds}
\label{sec:amplitudes}

If the $k=7$ torque channel is identified with an r-mode current quadrupole, the coefficient $\ell$ can be related to the dimensionless mode amplitude $\alpha$ through \citep{Owen1998}
\begin{equation}
\ell = \kappa_r \alpha^{2}, \qquad
\kappa_r =
\frac{\tilde{J}^{2}MR^{3}G(2\pi)^{4}}{5c^{5}},
\label{eq:kappa}
\end{equation}
where $\tilde{J}=1.635\times10^{-2}$ is the dimensionless current-quadrupole integral for the fiducial stellar model. This identification is an additional physical assumption: the timing formalism constrains the $k=7$ torque fraction, while the conversion to $\alpha$ assumes that this fraction is carried by an r-mode.

In the limiting case $f_7=1$, the observed spin-down is entirely attributed to the r-mode channel, giving the usual spin-down amplitude limit
\begin{equation}
\alpha_{\rm sd} =
\left(
\frac{5c^{5}|\dot{\Omega}|}
{8\pi G\Omega^{6}MR^{3}\tilde{J}^{2}}
\right)^{1/2},
\label{eq:alpha_sd}
\end{equation}
with $\Omega=2\pi\nu$.

The timing-based bounds on $f_7$ then propagate directly into bounds on $\alpha$. Using the capped physical upper bound from Eq.~(\ref{eq:f7max_phys}) gives
\begin{equation}
\alpha_{\max} =
\alpha_{\rm sd}\sqrt{f_{7,\max}^{\rm phys}},
\label{eq:amax}
\end{equation}
while the lower bound from Eq.~(\ref{eq:f7min}) yields
\begin{equation}
\alpha_{\min}
= \alpha_{\rm sd}\sqrt{f_{7,\min}}
=\alpha_{\rm sd}
\left[
\max\left(0,\frac{n_{\rm obs}-5}{2}\right)
\right]^{1/2}.
\label{eq:alpha_min}
\end{equation}

For sources in the formal domain $1\leq n_{\rm obs}\leq7$, Eq.~(\ref{eq:amax}) reduces to
\begin{equation}
\alpha_{\max}=
\alpha_{\rm sd}
\left(\frac{n_{\rm obs}-1}{6}\right)^{1/2}.
\label{eq:alpha_max}
\end{equation}

For transient effective braking indices outside this interval, the capped expression prevents the amplitude bound from being interpreted as a physical torque fraction larger than unity.

Equation~(\ref{eq:alpha_max}) provides a timing-based upper limit on the r-mode amplitude within the positive four-channel model, conditional on identifying the $k=7$ channel with an r-mode torque. By contrast, Eq.~(\ref{eq:alpha_min}) is nonzero only when $n_{\rm obs}>5$. Thus, for sources with $3<n_{\rm obs}<5$, timing may require an additional higher-order torque, but it does not by itself imply a nonzero lower bound on the r-mode amplitude. This distinction is important when interpreting sources such as J1640$-$4631 and J2229$+$6114.

The torque-fraction bounds themselves are independent of the neutron-star equation of state, but the conversion from $f_7$ to $\alpha$ and $h_0$ is not. From Eq.~(\ref{eq:alpha_sd}), the spin-down amplitude scales approximately as $\alpha_{\rm sd}\propto M^{-1/2}R^{-3/2}\tilde{J}^{-1}$ for fixed timing parameters. Therefore, changes in the stellar mass, radius, or current-quadrupole integral can shift the inferred amplitude by factors of a few. This stellar-structure uncertainty is larger than the formal uncertainty propagated from the measured braking index.

Table~\ref{tab:model_sensitivity} illustrates this sensitivity for the Crab pulsar: across a physically plausible range of masses and radii, the inferred amplitude varies by roughly a factor of four. Throughout this work, we therefore report central values using the fiducial model of Owen et al. (1998), with $M=1.4,M_{\odot}$ and $R=12.53,{\rm km}$, while emphasizing that any comparison with gravitational-wave upper limits must be made using a consistent stellar model.


\begin{table}[t]
\centering
\caption{Stellar-model sensitivity of the timing-based r-mode amplitude bound for the Crab pulsar. The calculation uses $f_{7,\max}^{\rm phys}=0.224$. The factor-of-few spread illustrates that the conversion from torque fraction to r-mode amplitude is dominated by stellar-structure systematics, while the torque-fraction bound itself is independent of the equation of state.}
\label{tab:model_sensitivity}
\footnotesize
\setlength{\tabcolsep}{5pt}
\renewcommand{\arraystretch}{1.15}
\begin{tabular}{cccc}
\hline
$M$ [$M_{\odot}$] & $R$ [km] & $\alpha_{\rm sd}$ & $\alpha_{\max}$ \tabularnewline
\hline
1.2 & 14.0  & $1.21\times10^{-5}$ & $5.72\times10^{-6}$ \tabularnewline
1.4 & 12.53 & $1.63\times10^{-5}$ & $7.70\times10^{-6}$ \tabularnewline
2.0 & 10.0  & $4.89\times10^{-5}$ & $2.31\times10^{-5}$ \tabularnewline
\hline
\end{tabular}
\end{table}

The corresponding gravitational-wave strain amplitude is
\begin{equation}
h_0=
\left(\frac{8\pi}{5}\right)^{1/2}
\frac{G}{c^4}
\frac{\alpha(2\pi f_{\rm gw})^2 M R^3 \tilde{J}}{d},
\label{eq:h0}
\end{equation}
where $f_{\rm gw}\approx(4/3)\nu$ and $d$ is the source distance \citep[Eq.~29]{LiAbbassi2025}. The canonical ratio $f_{\rm gw}=(4/3)\nu$ receives relativistic corrections of order a few percent for realistic stellar models \citep{AnderssonKokkotas2001}. These corrections are neglected in the present ranking-level estimates, but should be included in detailed continuous-wave searches.

Because $h_0$ depends on the same stellar parameters as the amplitude conversion, the strain estimates inherit the same equation-of-state dependence. Consequently, a gravitational-wave upper limit should be compared with the timing-based prediction only after adopting a common stellar model and distance estimate.

\begin{figure}[!ht]
\centering
\includegraphics[width=\columnwidth]{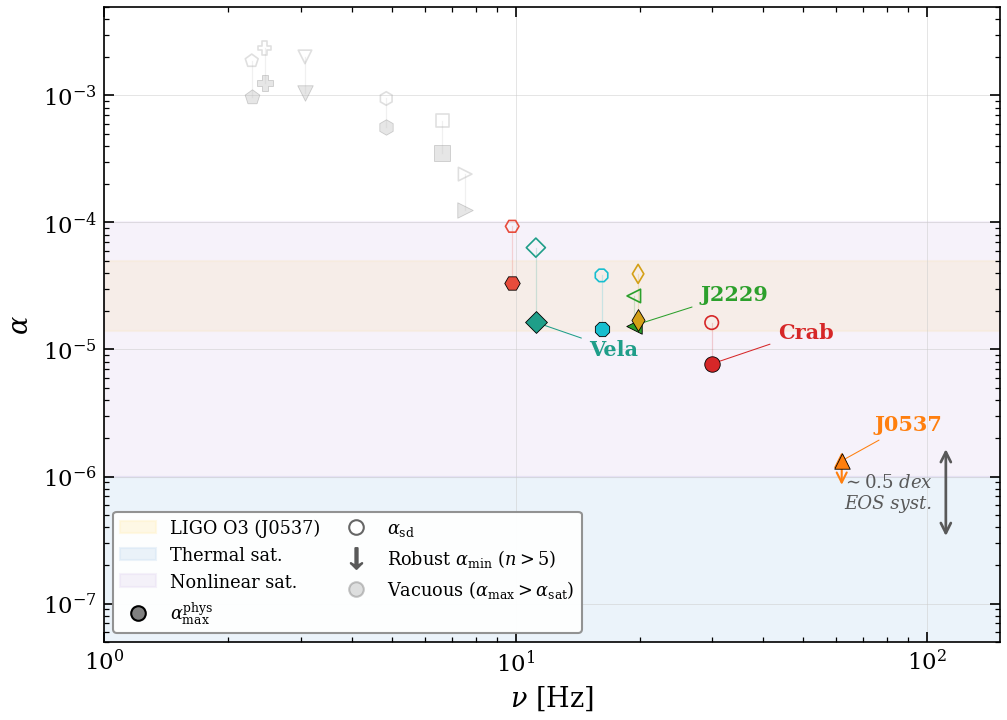}
\caption{Timing-based r-mode amplitude constraints as a
function of spin frequency. Filled markers show the upper
bounds $\alpha_{\max}$ obtained from the capped physical torque
fraction $f_{7,\max}^{\rm phys}$, while open markers show
the full spin-down limit $\alpha_{\rm sd}$ corresponding to $f_7=1$.
Sources with $\alpha_{\max}>\alpha_{\rm sat}\sim10^{-4}$, for which
the timing bound is above the illustrative saturation scale,
are shown in grey. The robust lower bound $\alpha_{\min}$ is
nonzero only for $n_{\rm obs}>5$, which in the present formal
sample applies exclusively to J0537. Shaded bands indicate
the LIGO~O3 limit for J0537 together with illustrative
theoretical saturation ranges. All amplitudes carry an
EOS-dependent systematic uncertainty of order
$\sim0.5$~dex, as illustrated in
Table~\ref{tab:model_sensitivity}.}
\label{fig:alpha_bounds}
\end{figure}

Figure~\ref{fig:alpha_bounds} presents the resulting amplitude census. The timing data provide upper limits for the full formal sample, while a strictly nonzero lower amplitude bound appears only for sources in the $n_{\rm obs}>5$ regime. PSR~J0537$-$6910 remains the most interesting case: its near-asymptotic inter-glitch braking index approaches $n\simeq7$, so that the upper and lower bounds converge toward $\alpha_{\max}\simeq\alpha_{\min}\simeq1.3\times10^{-6}$ under the fiducial stellar model. This places the inferred amplitude within the broad range discussed for CFS-driven saturation and below the current LIGO~O3 upper limit when the same stellar model is used. The interpretation remains conditional, however, because the inter-glitch braking behaviour may also arise from superfluid-recovery dynamics rather than from an r-mode torque.

\subsection{Corrected ages}
\label{sec:ages}

For a constant effective braking index $n_{\rm obs}$, the usual
characteristic-age estimate generalizes to
\begin{equation}
\tau_n
=
\frac{\nu}{(n_{\rm obs}-1)|\dot{\nu}|}
=
\frac{2\tau_c}{n_{\rm obs}-1},
\label{eq:tau_n}
\end{equation}
where $\tau_c=\nu/(2|\dot{\nu}|)$ is the standard characteristic
age. This expression is the natural constant-$n$ analogue of
$\tau_c$, obtained under the same implicit assumption that the
birth spin frequency was substantially larger than the present
one.

If the birth spin frequency $\nu_0$ is not negligible, the
corresponding constant-$n$ spin-down age becomes
\begin{equation}
t
=
\frac{\nu}{(n_{\rm obs}-1)|\dot{\nu}|}
\left[
1-
\left(
\frac{\nu}{\nu_0}
\right)^{n_{\rm obs}-1}
\right].
\label{eq:tau_birth}
\end{equation}
Thus, Eq.~(\ref{eq:tau_n}) should be interpreted as the
large-birth-spin limiting form of the constant-$n$ age, rather
than as a unique historical age. This distinction is important
because the torque-fraction bounds derived in
Section~\ref{sec:bounds} depend only on the present measured
braking index and are insensitive to the assumed initial spin
frequency, whereas any chronological interpretation of $\tau_n$
is not.

For Vela, with $n_{\rm obs}=1.4$, Eq.~(\ref{eq:tau_n}) gives
$\tau_n\simeq57\,{\rm kyr}$, compared with
$\tau_c\simeq11.4\,{\rm kyr}$. For PSR~J0537$-$6910, with
$n_{\rm obs}\simeq7$, it gives
$\tau_n\simeq1.6\,{\rm kyr}$, compared with
$\tau_c\simeq4.9\,{\rm kyr}$.

These corrected ages should not be confused with independently
inferred supernova-remnant or pulsar-wind-nebula ages. A fully
uniform source-by-source comparison is limited because reliable
external ages are not available for all pulsars in the timing
sample. Vela provides the clearest example: its present
braking-index-corrected value,
$\tau_n\simeq57\,{\rm kyr}$, is substantially larger than the
commonly adopted age of the Vela supernova remnant. This
difference indicates that the currently measured braking index is
unlikely to represent the lifetime-averaged spin-down history.
Comparable discrepancies should be interpreted as evidence for
torque evolution, glitch recovery, changing magnetospheric
conditions, or a non-negligible birth spin, rather than as failures
of the instantaneous torque-fraction bracket.

Equation~(\ref{eq:tau_n}) also assumes that the effective braking
index has remained constant throughout the secular spin-down
history, which is unlikely to hold exactly in real systems.
Accordingly, $\tau_n$ should not be interpreted as an exact
chronometer. It is more appropriately regarded as a
first-order diagnostic of the present effective torque law. This
qualification is particularly important for sources such as Vela,
whose measured $n_{\rm obs}=1.4$ likely reflects the current
torque balance rather than a lifetime-averaged braking law.

\begin{figure}[!ht]
\centering
\includegraphics[width=\columnwidth]{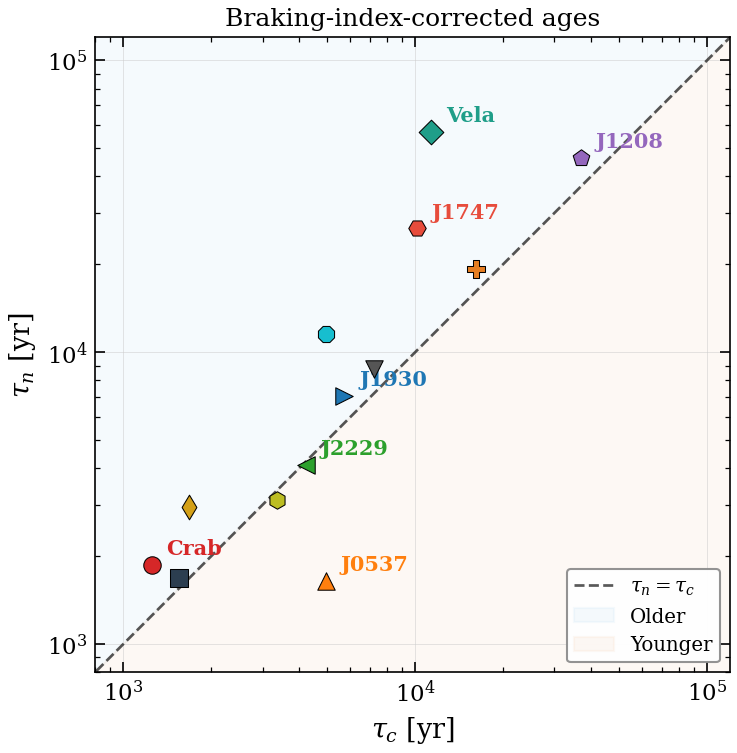}
\caption{Braking-index-corrected ages $\tau_n$ versus the standard
characteristic ages $\tau_c$. The dashed diagonal marks
$\tau_n=\tau_c$. Sources above the line have
$\tau_n>\tau_c$, whereas sources below it have
$\tau_n<\tau_c$. The large displacement of Vela illustrates
that the present braking index need not represent the
lifetime-averaged spin-down history.}
\label{fig:corrected_ages}
\end{figure}

Figure~\ref{fig:corrected_ages} presents the
braking-index-corrected ages in the $(\tau_c,\tau_n)$ plane.
Sources with $n_{\rm obs}<3$ have $\tau_n>\tau_c$ and therefore
appear older than the standard dipole estimate, whereas sources
with $n_{\rm obs}>3$ have $\tau_n<\tau_c$. The most extreme
case is Vela, with $\tau_n/\tau_c\simeq5$. Its displacement from
the diagonal should be interpreted as further evidence that the
present braking index is not a reliable lifetime-averaged index,
rather than as a precise revision of the true Vela age.

\begin{table*}[t]
\centering

\caption{Timing constraints on the $k=7$ torque contribution and
the corresponding r-mode amplitude bounds for the young-pulsar
sample. Panel (a) lists the measured spin frequencies and braking
indices together with the allowed torque-fraction interval in the
positive four-channel secular model. Panel (b) gives the associated
r-mode amplitude limits, braking-index-corrected ages, and idealized
Einstein Telescope ranking index. The formal analysis includes the
13 sources satisfying $1\leq n_{\rm obs}\leq7$; PSR~J1734$-$3333
is excluded because its sub-unity braking index lies outside this
domain. Calculations adopt the fiducial stellar model
$M=1.4\,M_{\odot}$ and $R=12.53\,{\rm km}$. The symbol
$\dagger$ indicates $\alpha_{\max}>\alpha_{\rm sat}\sim10^{-4}$,
and $\ddagger$ denotes the pre-outburst timing state of
PSR~J1846$-$0258. For PSR~J0537$-$6910, the near-asymptotic
inter-glitch braking index is used. Quoted errors are
$1\sigma$ observational uncertainties and are propagated into
the derived torque-fraction and amplitude bounds where reported.}

\label{tab:results}

\scriptsize
\setlength{\tabcolsep}{3.4pt}
\renewcommand{\arraystretch}{1.13}

\noindent\textit{(a) Timing parameters and allowed $k=7$ torque fractions}

\medskip

\begin{tabular}{lccccc}
\hline
Source &
$\nu$ [Hz] &
$n_{\rm obs}$ &
$f_{7,\min}$ &
$f_{7,\max}^{\rm phys}$ &
Class
\tabularnewline
\hline

Crab (B0531+21)
& 29.95
& $2.342\pm0.001$
& $0$
& $0.2237\pm0.0002$
& EM
\tabularnewline

B1509$-$58
& 6.633
& $0$
& $0.3065\pm0.0002$
& EM
\tabularnewline

Vela (B0833$-$45)
& 11.19
& $1.40\pm0.20$
& $0$
& $0.067\pm0.033$
& EM
\tabularnewline

J1119$-$6127
& 2.451
& $2.684\pm0.002$
& $0$
& $0.2807\pm0.0003$
& EM
\tabularnewline

J1846$-$0258$\ddagger$
& 3.073
& $2.65\pm0.01$
& $0$
& EM
\tabularnewline

J1640$-$4631
& 4.843
& $3.15\pm0.03$
& $0$
& $0.358\pm0.005$
& HO
\tabularnewline

J0537$-$6910 (ig)
& 62.00
& $7.0\pm0.3$
& $1.00^{+0.00}_{-0.15}$
& $1.00^{+0.00}_{-0.05}$
& $f_7>0$
\tabularnewline

J1208$-$6238
& 2.281
& $2.598\pm0.001_{\rm stat}\pm0.100_{\rm sys}$
& $0$
& $0.266\pm0.017$
& EM
\tabularnewline

J1747$-$2958
& 9.803
& $1.766$
& $0$
& $0.128$
& EM
\tabularnewline

J1833$-$1034
& 16.16
& $1.8569\pm0.0006$
& $0$
& $0.1428\pm0.0001$
& EM
\tabularnewline

J1930$+$1852
& 7.532
& $2.613$
& $0$
& $0.269$
& EM
\tabularnewline

J2229$+$6114
& 19.36
& $3.050$
& $0$
& $0.342$
& HO
\tabularnewline

B0540$-$69
& 19.83
& $2.140\pm0.009$
& $0$
& $0.190\pm0.002$
& EM
\tabularnewline

\hline
\end{tabular}

\bigskip

\noindent\textit{(b) Amplitude limits, corrected ages, and ET ranking}

\medskip

\begin{tabular}{lccccccc}
\hline
Source &
$\alpha_{\rm sd}$ &
$\alpha_{\min}$ &
$\alpha_{\max}$ &
$\alpha_{\max}/\alpha_{\rm sd}$ &
$\tau_c$ [kyr] &
$\tau_n$ [kyr] &
$\mathcal{D}_{\rm ET}^{\rm ideal}$
\tabularnewline
\hline

Crab (B0531+21)
& $1.6{\times}10^{-5}$
& --
& $(7.70\pm0.003){\times}10^{-6}$
& 0.47
& 1.25
& 1.87
& 26.0
\tabularnewline

B1509$-$58
& $6.3{\times}10^{-4}$
& --
& $(3.500\pm0.001){\times}10^{-4}\dagger$
& 0.55
& 1.55
& 1.69
& 10.5
\tabularnewline

Vela (B0833$-$45)
& $6.3{\times}10^{-5}$
& --
& $(1.6\pm0.4){\times}10^{-5}$
& 0.26
& 11.4
& 56.8
& 32.8
\tabularnewline

J1119$-$6127
& $2.4{\times}10^{-3}$
& --
& $(1.3000\pm0.0008){\times}10^{-3}\dagger$
& 0.53
& 16.2
& 19.2
& 1.9
\tabularnewline

J1846$-$0258$\ddagger$
& $2.0{\times}10^{-3}$
& --
& $(1.100\pm0.003){\times}10^{-3}\dagger$
& 0.52
& 7.2
& 8.8
& 4.0
\tabularnewline

J1640$-$4631
& $9.5{\times}10^{-4}$
& --
& $(5.70\pm0.04){\times}10^{-4}\dagger$
& 0.60
& 3.4
& 3.1
& 3.3
\tabularnewline

J0537$-$6910 (ig)
& $1.3{\times}10^{-6}$
& $1.30^{+0.00}_{-0.10}{\times}10^{-6}$
& $1.30^{+0.00}_{-0.03}{\times}10^{-6}$
& 1.00
& 4.9
& 1.63
& 1.10
\tabularnewline

J1208$-$6238
& $1.9{\times}10^{-3}$
& --
& $(0.970\pm0.030){\times}10^{-3}\dagger$
& 0.52
& 37.0
& 46.3
& 3.7
\tabularnewline

J1747$-$2958
& $9.3{\times}10^{-5}$
& --
& $3.3{\times}10^{-5}$
& 0.36
& 10.2
& 26.6
& 2.7
\tabularnewline

J1833$-$1034
& $3.8{\times}10^{-5}$
& --
& $(1.5000\pm0.0005){\times}10^{-5}$
& 0.38
& 4.9
& 11.5
& 5.1
\tabularnewline

J1930$+$1852
& $2.4{\times}10^{-4}$
& --
& $1.3{\times}10^{-4}\dagger$
& 0.52
& 5.7
& 7.1
& 3.8
\tabularnewline

J2229$+$6114
& $2.6{\times}10^{-5}$
& --
& $1.5{\times}10^{-5}$
& 0.58
& 4.2
& 4.1
& 11.7
\tabularnewline

B0540$-$69
& $3.9{\times}10^{-5}$
& --
& $(1.700\pm0.007){\times}10^{-5}$
& 0.44
& 1.7
& 3.0
& 0.8
\tabularnewline

\hline
\end{tabular}

\medskip

\begin{minipage}{0.98\textwidth}
\footnotesize
\textit{Notes.}
EM denotes sources whose braking indices are consistent with the
wind-plus-dipole sector of the positive four-channel model. HO
denotes sources requiring an additional higher-order torque,
although the $k=7$ fraction may still vanish. The class $f_7>0$
indicates a strictly positive lower bound on the $k=7$ contribution.
Within rounding,
$\alpha_{\max}/\alpha_{\rm sd}
=\left(f_{7,\max}^{\rm phys}\right)^{1/2}$.
The amplitude bounds assume the fiducial stellar model and carry
stellar-structure uncertainties of order a factor of a few.

Where uncertainties in $n_{\rm obs}$ are reported, they are
propagated using $\sigma_{f_{7,\max}}=\sigma_n/6$ and, for
$n_{\rm obs}>5$, $\sigma_{f_{7,\min}}=\sigma_n/2$, with
$\sigma_{\alpha}=\alpha_{\rm sd}\sigma_f/(2\sqrt{f})$. The physical
restriction $0\leq f_7\leq1$ produces asymmetric uncertainties for
PSR~J0537$-$6910, while the uncertainty for PSR~J1208-6238 is
dominated by the reported systematic term. For PSRs~J1747$-$2958$,
J1930$+$1852$, and J2229$+$6114, no measurement uncertainties were
reported in the adopted timing references. The propagated errors do
not alter the source classifications or the principal conclusions.
\end{minipage}

\end{table*}

\subsection{Gravitational-wave ranking}
\label{sec:gw_ranking}

We define the idealized Einstein Telescope ranking metric as
\begin{equation}
\mathcal{D}_{\rm ET}^{\rm ideal}
\equiv
\frac{h_0^{\max}}
     {h_{\rm ET}^{1\,{\rm yr}}(f_{\rm gw})},
\label{eq:DET_ideal}
\end{equation}
where $h_0^{\max}$ is the timing-based upper bound on the
gravitational-wave strain and
$h_{\rm ET}^{1\,{\rm yr}}(f_{\rm gw})$ is the nominal one-year
coherent ET-D strain sensitivity evaluated at the corresponding
gravitational-wave frequency. Thus,
$\mathcal{D}_{\rm ET}^{\rm ideal}>1$ indicates that the
timing-based upper bound lies above the nominal ET-D sensitivity,
whereas $\mathcal{D}_{\rm ET}^{\rm ideal}<1$ indicates that it lies
below that sensitivity. This dimensionless quantity is intended as
a comparative source-ranking metric rather than as a full
detection forecast.

\begin{figure}[!ht]
\centering
\includegraphics[width=\columnwidth]{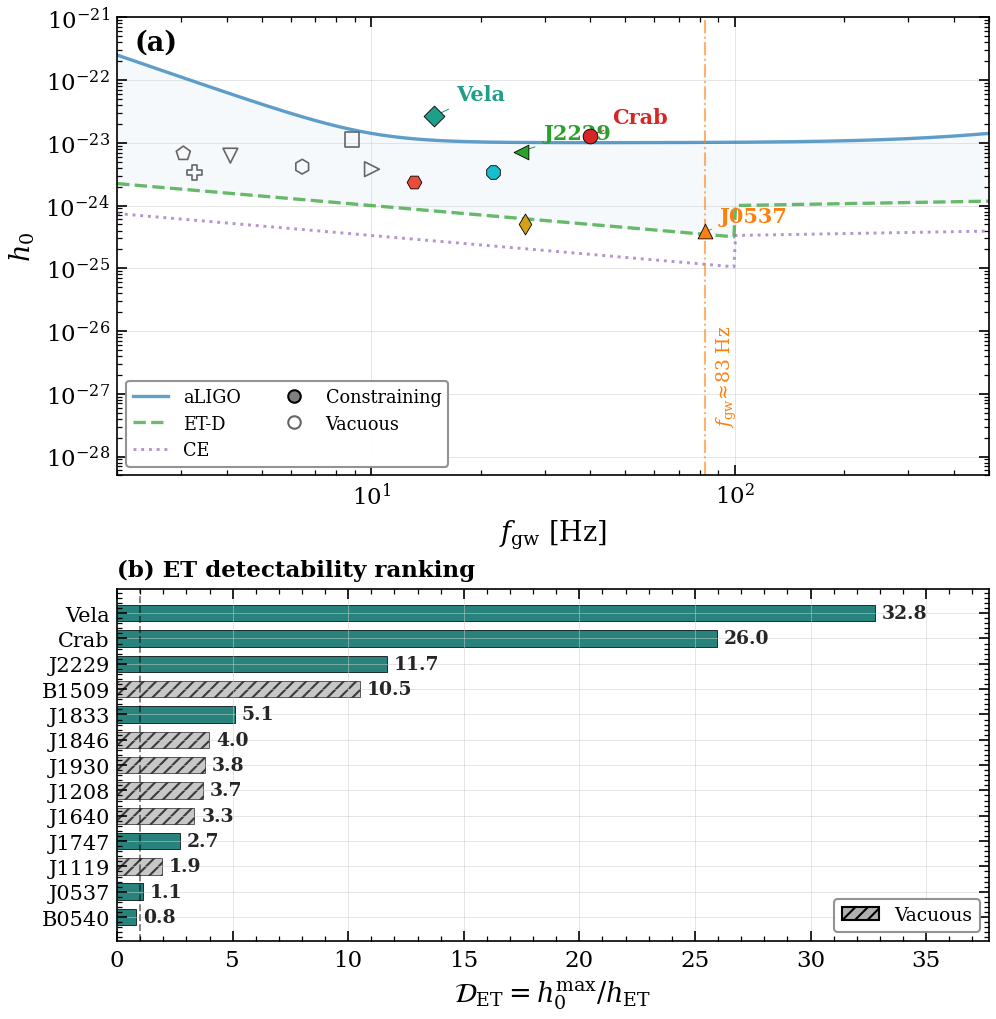}
\caption{(a) Timing-based strain upper limits $h_0^{\max}$ as a
function of $f_{\rm gw}$, shown together with representative
sensitivity curves for Advanced LIGO, ET-D, and Cosmic Explorer.
The vertical marker at $f_{\rm gw}\simeq83\,{\rm Hz}$ indicates
the expected r-mode signal frequency for PSR~J0537$-$6910.
Filled and open symbols distinguish constraining and vacuous
timing-based bounds under the adopted saturation benchmark.
(b) Idealized ET-D ranking metric
$\mathcal{D}_{\rm ET}^{\rm ideal}$, ordered by source. This
quantity is intended only as a comparative guide and should not
be interpreted as a full detection forecast. The corresponding
optimistic 30-day coherence-limited values are listed in
Table~\ref{tab:DET30d}.}
\label{fig:ET_forecast}
\end{figure}

Figure~\ref{fig:ET_forecast} shows the corresponding
gravitational-wave ranking. Under the adopted stellar model and
an idealized one-year coherent ET-D comparison, the largest values
are obtained for Vela,
$\mathcal{D}_{\rm ET}^{\rm ideal}\simeq32.8$, the Crab,
$\mathcal{D}_{\rm ET}^{\rm ideal}\simeq26.0$, and
PSR~J2229$+$6114,
$\mathcal{D}_{\rm ET}^{\rm ideal}\simeq11.7$. These sources are
therefore the most constraining cases for future non-detections
under the timing-based upper limits. PSR~J0537$-$6910, with
$\mathcal{D}_{\rm ET}^{\rm ideal}\simeq1.1$, lies closest to the
nominal ET-D threshold and remains the most physically
interesting prospective detection target if its higher-order
torque is genuinely associated with an r-mode.

The idealized ranking assumes a year-long coherent integration
and neglects the practical effects of glitches, parameter-space
trials, template mismatch, non-stationary detector noise, and
duty-cycle losses. These limitations are especially relevant for
PSR~J0537$-$6910 because of its frequent glitches.

To illustrate the effect of a shorter glitch-limited
coherence interval, we define
\begin{equation}
\mathcal{D}_{\rm ET}^{30{\rm d}}
=
\mathcal{D}_{\rm ET}^{\rm ideal}
\left(
\frac{30\,{\rm d}}
     {365\,{\rm d}}
\right)^{1/2}
\simeq
0.287\,\mathcal{D}_{\rm ET}^{\rm ideal}.
\label{eq:DET_30d}
\end{equation}
The square-root dependence in
Eq.~(\ref{eq:DET_30d}) corresponds to a fully coherent
30-day integration. It should therefore be interpreted as an
optimistic coherence-limited estimate, not as the sensitivity of a
realistic semi-coherent search. In practice, incoherent
combination losses and additional search inefficiencies generally
lead to a weaker sensitivity scaling.

Table~\ref{tab:DET30d} lists the resulting 30-day
coherence-limited ranking values for the full sample.

\begin{table}[!ht]
\centering
\caption{Idealized one-year and 30-day ET-D ranking metrics for the
young-pulsar sample. The 30-day values are obtained from
Eq.~(\ref{eq:DET_30d}). They represent optimistic
coherence-limited estimates for a fully coherent 30-day segment
and should not be interpreted as realistic semi-coherent search
sensitivities.}
\label{tab:DET30d}
\footnotesize
\setlength{\tabcolsep}{5pt}
\renewcommand{\arraystretch}{1.12}

\begin{tabular}{lcc}
\hline
Source &
$\mathcal{D}_{\rm ET}^{\rm ideal}$ &
$\mathcal{D}_{\rm ET}^{30{\rm d}}$
\tabularnewline
\hline

Crab (B0531+21)   & 26.0 & 7.46 \tabularnewline
B1509$-$58        & 10.5 & 3.01 \tabularnewline
Vela (B0833$-$45) & 32.8 & 9.41 \tabularnewline
J1119$-$6127      & 1.9  & 0.55 \tabularnewline
J1846$-$0258      & 4.0  & 1.15 \tabularnewline
J1640$-$4631      & 3.3  & 0.95 \tabularnewline
J0537$-$6910 (ig) & 1.1  & 0.32 \tabularnewline
J1208$-$6238      & 3.7  & 1.06 \tabularnewline
J1747$-$2958      & 2.7  & 0.77 \tabularnewline
J1833$-$1034      & 5.1  & 1.46 \tabularnewline
J1930$+$1852      & 3.8  & 1.09 \tabularnewline
J2229$+$6114      & 11.7 & 3.36 \tabularnewline
B0540$-$69        & 0.8  & 0.23 \tabularnewline

\hline
\end{tabular}
\end{table}

With this scaling, the ranking of PSR~J0537$-$6910 decreases
from
$\mathcal{D}_{\rm ET}^{\rm ideal}\simeq1.1$
to
$\mathcal{D}_{\rm ET}^{30{\rm d}}\simeq0.32$.
Thus, J0537 remains the cleanest physical test case for the
r-mode interpretation, but a realistic glitch-limited ET search
would not automatically guarantee detection at the fiducial
amplitude. The corresponding 30-day values for Vela, the Crab,
and PSR~J2229$+$6114 are approximately $9.4$, $7.5$, and
$3.4$, respectively.

The ranking separates two related but distinct questions.
Sources such as Vela and the Crab are especially useful for
placing constraining upper limits because their timing-based
strain bounds are comparatively high relative to the assumed
detector sensitivity. PSR~J0537$-$6910, by contrast, is the most
physically interesting r-mode test case because its inter-glitch
braking index approaches the $k=7$ regime. The broader
astrophysical plausibility of these interpretations is considered
below through the instability-window analysis.

\section{Discussion}
\label{sec:discussion}

\subsection{GW searches}

The timing-based framework developed here and direct gravitational-wave searches probe the same physical question from different but complementary directions. Continuous-wave searches constrain the instantaneous strain amplitude $h_0$, whereas the present method constrains the fractional contribution of a putative $k=7$ current-quadrupole channel to the secular spin-down. In this sense, timing and gravitational-wave data do not compete; rather, they test different projections of the same underlying torque budget.

For PSR~J0537$-$6910, the main observational degeneracy is between a genuine r-mode current-quadrupole torque and a non-radiative vortex-creep or superfluid-recovery interpretation of the inter-glitch braking index. In the r-mode interpretation, the large inter-glitch value $n_{\rm ig}\simeq7$ is attributed to a quasi-steady $k=7$ torque, implying $\alpha\approx1.3\times10^{-6}$ under the fiducial stellar model, a continuous-wave signal near $f_{\rm gw}\approx 83,{\rm Hz}$, and a fiducial strain $h_{0,\rm fid}\approx3.8\times10^{-25}$. In the vortex-creep interpretation, the same effective braking behaviour is produced by time-dependent post-glitch recovery in $\ddot{\nu}$, with no required r-mode amplitude, no required $h_0$, and no mandatory continuous-wave signal. Thus, a combination of glitch-resolved timing, the evolution of $\ddot{\nu}$ between glitches, and continuous-wave upper limits provides the clearest way to break this degeneracy.

For PSR~J0537$-$6910, the timing-based upper bound $\alpha_{\max}\approx1.3\times10^{-6}$ is formally more restrictive than the current LIGO O3 bound, approximately $\alpha<1.4\times10^{-5}$--$5\times10^{-5}$ \citep{Abbott2021_J0537_rmode}, provided both are evaluated using the same stellar model. This comparison, however, should be interpreted carefully. The timing bound is indirect and model dependent: it follows from the positive four-channel secular spin-down framework, from the capped physical torque fraction $f_{7,\max}^{\rm phys}$, and from the assumption that the relevant inter-glitch behaviour is tracing a quasi-steady torque balance. By contrast, the LIGO limit is direct, but it applies to the instantaneous gravitational-wave signal and is affected by the practical difficulty of searching a frequently glitching source.

A future detection near $f_{\rm gw}\approx83,{\rm Hz}$ from J0537 would be especially powerful. Through Eq.~(\ref{eq:h0}), it would provide a direct measurement of $\alpha$, which could then be combined with the timing-based bracket on $f_7$ and the stellar-structure assumptions used to convert torque fraction into amplitude and strain. Conversely, a sufficiently deep non-detection would place pressure on the simplest r-mode interpretation of the
inter-glitch timing behaviour.

For the fiducial stellar model used here, the corresponding timing-based strain is $h_{0,\rm fid}\approx3.8\times10^{-25}$. A 95\% upper limit below this level, obtained with a phase model that accounts for glitches, would rule out the simplest fiducial r-mode-dominated interpretation of J0537.

This statement is conditional because the predicted strain depends on the stellar mass, radius, current-quadrupole normalization, distance, and the assumed coherence of the signal across glitches. A year-long coherent search is optimistic for J0537 because of its frequent glitches. A realistic semi-coherent or glitch-limited search may therefore be less sensitive than the idealized ranking used here.

The broader value of the present framework is therefore not that it turns timing data into a unique gravitational-wave prediction. Rather, it identifies which sources are most worth targeting, separates direct gravitational-wave constraints from timing-based torque constraints, and clarifies what combination of timing and continuous-wave observations would support or disfavor an r-mode interpretation.

\subsection{Instability window}
\label{sec:cfs_window}

A key interpretive step remains: even when timing permits or
suggests a higher-order torque contribution, that does not by
itself guarantee that the contribution should be identified with
an r-mode. For such an interpretation to be physically credible,
the star must also lie in a region of parameter space where the
Chandrasekhar--Friedman--Schutz instability is expected to operate.

Classical calculations \citep{AnderssonKokkotas2001,Owen1998}
showed that young, rapidly rotating neutron stars can enter the
r-mode instability window. More recent work has demonstrated that
the extent of the unstable region is highly sensitive to the
assumed microphysics. Enhanced bulk viscosity in superfluid matter
can substantially reduce the unstable region
\citep{GusakovChugunov2019}, resonant mutual friction can suppress
the instability further \citep{Kraav2024}, and nonlinear
saturation mechanisms may restrict the mode amplitude to very
small values in some temperature--frequency regimes
\citep{AlfordSchwenzer2014a}. It is therefore important to
distinguish between two separate questions: whether timing permits
an r-mode contribution, and whether such an interpretation is
astrophysically plausible for a particular source.

\begin{figure}[!ht]
\centering
\includegraphics[width=\columnwidth]{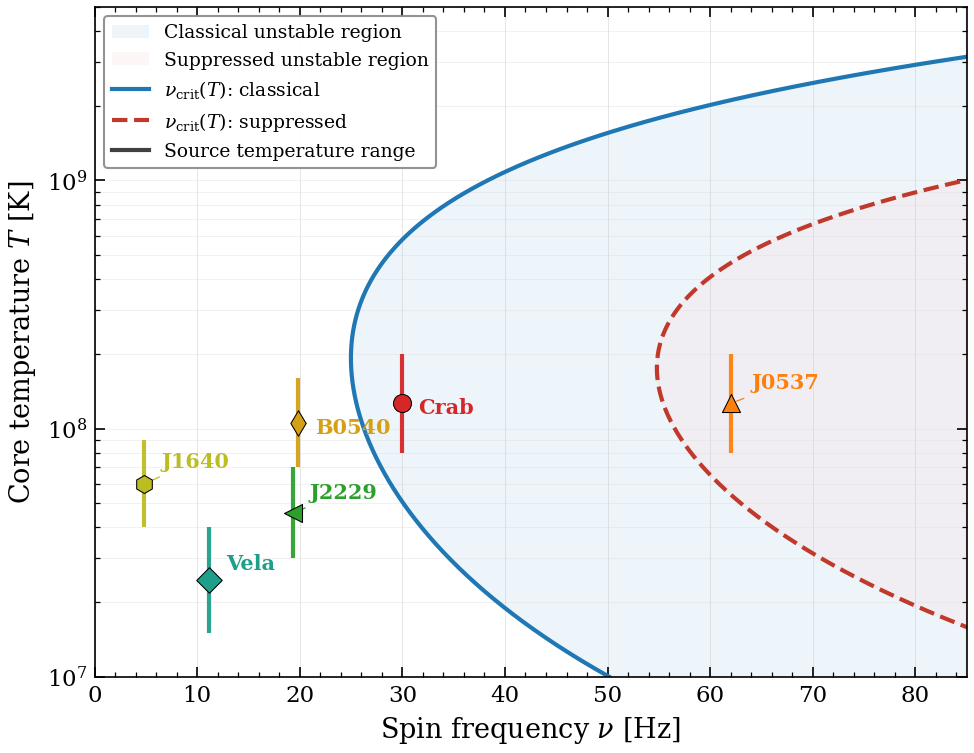}
\caption{Schematic r-mode instability window in the
spin-frequency--temperature plane. The broader shaded region
represents a classical unstable window, whereas the narrower
region illustrates the possible effect of stronger microphysical
damping, including enhanced bulk viscosity and mutual friction.
For the Crab and PSR~J0537$-$6910, the vertical ranges
represent approximate core temperatures inferred from
representative cooling calculations:
$T_{\rm core}^{\rm Crab}\sim(1\text{--}3)\times10^{8}\,{\rm K}$
and
$T_{\rm core}^{\rm J0537}\sim(0.5\text{--}2)\times10^{8}\,{\rm K}$.
These are not direct temperature measurements. Enhanced neutrino
cooling could place either source at temperatures of a few
$\times10^{7}\,{\rm K}$.} The source locations should therefore
be interpreted qualitatively, because their inferred thermal
states depend on uncertain cooling physics, composition, pairing,
and age. The figure illustrates that an r-mode interpretation is
most plausible for young and comparatively rapidly rotating
sources such as PSR~J0537$-$6910 and, more tentatively, the Crab,
whereas older and cooler pulsars are more likely to lie near or
below the stability boundary.
\label{fig:instability_window}
\end{figure}

Figure~\ref{fig:instability_window} shows a schematic instability
window in the $(\nu,T)$ plane. The figure is intended as an
interpretive guide rather than as a detailed source-by-source
microphysical classification. It illustrates how the plausibility
of an r-mode interpretation depends jointly on spin frequency,
thermal state, and the assumed strength of viscous damping.

The core temperatures of the Crab and
PSR~J0537$-$6910 have not been measured directly. Approximate
internal-temperature ranges can nevertheless be estimated by
comparing their ages with representative neutron-star cooling
calculations. Such estimates are intrinsically model dependent and
vary with stellar mass, composition, nucleon pairing, envelope
properties, and the adopted neutrino-cooling mechanism
\citep{YakovlevPethick2004,Potekhin2015}.

For the Crab, whose historical age is approximately
$1\,{\rm kyr}$, standard or minimal-cooling models typically give
\[
\mathbf{
T_{\rm core}^{\rm Crab}
\sim
(1\text{--}3)\times10^{8}\,{\rm K}.
}
\]
For PSR~J0537$-$6910, adopting
$\tau_c\simeq4.9\,{\rm kyr}$ as an approximate age scale gives
\[
\mathbf{
T_{\rm core}^{\rm J0537}
\sim
(0.5\text{--}2)\times10^{8}\,{\rm K}.
}
\]

These intervals should be interpreted as representative
cooling-model estimates rather than observational constraints.
Their widths reflect uncertainties in stellar structure,
composition, superfluid pairing, and the heat-blanketing envelope.
Moreover, enhanced neutrino cooling, including direct-Urca
emission in a sufficiently massive neutron star, could reduce the
core temperature to a few $\times10^{7}\,{\rm K}$ at comparable
ages \citep{YakovlevPethick2004,Potekhin2015}. The thermal
locations of the Crab and J0537 in
Figure~\ref{fig:instability_window} are therefore shown as broad
temperature intervals rather than as precisely determined
points.

Within this framework, the fastest and youngest objects in the
sample remain the most plausible r-mode candidates. In particular,
PSR~J0537$-$6910, with $\nu=62\,{\rm Hz}$, and the Crab, with
$\nu\simeq30\,{\rm Hz}$, are the sources most naturally placed
within or near an unstable region if their core temperatures remain
of order $10^{8}\,{\rm K}$. The representative ranges above show
that such temperatures are plausible under standard or minimal
cooling, but they are not guaranteed. Enhanced neutrino cooling
could move either source below the nominal instability boundary,
and their classification relative to the instability window
therefore remains cooling-model dependent.

By contrast, older and more slowly rotating systems such as Vela,
PSR~J1747$-$2958, and PSR~J1208$-$6238 are increasingly likely to
have cooled below the instability boundary, particularly if
enhanced cooling or stronger viscous damping is operative.

This perspective also clarifies the interpretation of sources with
intermediate braking indices. PSR~J1640$-$4631 and
PSR~J2229$+$6114 require a higher-order contribution beyond the
wind-plus-dipole electromagnetic sector, but timing alone does not
require that contribution to be an r-mode torque. Their placement
in spin-frequency--temperature space likewise does not make the
r-mode interpretation unique. Within the present sample, the
strongest combined timing-based and astrophysical case for a
possible r-mode contribution therefore remains
PSR~J0537$-$6910, with the Crab representing a more tentative
secondary candidate.

\subsection{Limitations}
\label{sec:limitations}

The framework developed in this paper is intentionally phenomenological, and its usefulness depends on respecting the limits of that interpretation. The analysis assumes that the effective torque coefficients are non-negative and vary slowly over the timescale on which the braking index is measured. In real pulsars, however, glitch activity, magnetospheric state changes, secular magnetic-field evolution, and internal coupling processes can all introduce time dependence. When such effects are important, the coefficients in the model should be understood as effective secular parameters rather than direct microscopic torque amplitudes, or the source may fall outside the domain of the steady four-channel interpretation.

A further limitation is the assumption that the spin-down law can be represented by integer odd powers of $\nu$. While this is a natural first approximation under rotation-reversal antisymmetry, more complicated plasma or superfluid dynamics may lead to departures from this structure. \citet{Chishtie2026}, for example, argued that symmetry breaking in pulsar spin-down phenomenology can generate non-integer effective exponents. In such cases, the present formalism remains useful as a lowest-order diagnostic, but it should not be regarded as fully general.

It is also important to distinguish between algebraic robustness and physical uniqueness. The timing brackets derived here are clean consequences of the positive four-channel model. By contrast, identifying a higher-order torque with a specific physical mechanism requires additional input. For sources with $3<n_{\rm obs}<5$, timing alone shows that a torque beyond the electromagnetic sector is required, but it does not uniquely isolate the r-mode channel. Even for J0537, where the timing evidence is most suggestive, viable superfluid-dynamical alternatives such as vortex-creep recovery remain.

Finally, the torque-fraction bounds themselves are independent of the neutron-star equation of state, but the conversion to $\alpha$, $h_0$, and detectability is not. These quantities inherit uncertainties from the stellar mass, radius, current-quadrupole normalization, distance, and thermal state. These systematics are larger than the formal uncertainties propagated from the measured braking indices and should be included in any comparison with gravitational-wave upper limits.

Despite these limitations, the framework offers a useful and systematic first step. It extracts nontrivial information from timing data alone, distinguishes robust algebraic conclusions from model-dependent physical interpretations, and helps identify the sources for which joint timing and gravitational-wave analysis is most promising. Its main value is not that it provides a final answer to the r-mode problem, but that it supplies a transparent observational diagnostic with clear paths to falsification.

\section{Conclusions}
\label{sec:conclusions}

We have developed and applied a timing-based framework for constraining higher-order spin-down torques in young pulsars through braking-index measurements. The main results are as follows.

\begin{enumerate}

\item Within the positive four-channel model, the braking index is a torque-weighted average of the exponents ${1,3,5,7}$. Positivity of the torque fractions gives the timing bracket $f_{7,\min}=\max[0,(n_{\rm obs}-5)/2]$ and $f_{7,\max}^{\rm phys}=\min{1,\max[0,(n_{\rm obs}-1)/6]}$. Sources with $n_{\rm obs}>3$ require a higher-order torque beyond the wind-plus-dipole electromagnetic sector, but only sources with $n_{\rm obs}>5$ require a strictly nonzero $k=7$ contribution within the full positive four-channel model. This distinction is important because a braking index above the dipole value does not by itself imply r-mode spin-down.

\item In the restricted gravitational-wave sector, the threshold $n=5$ separates mass-quadrupole-like and current-quadrupole-like dominance. Within the formal census, only PSR~J0537$-$6910 lies above this boundary. This inference is conditional on the assumption that the effective torque coefficients are non-negative and slowly varying over the timing interval. If rapidly varying glitch recovery, magnetospheric switching, magnetic-field evolution, or other time-dependent effects dominate the measured braking index, the torque-fraction interpretation need not apply.

\item The generalized characteristic-age estimate $\tau_n=2\tau_c/(n_{\rm obs}-1)$ revises the standard dipole age by factors ranging from $\sim5$ for Vela to $\sim0.33$ for J0537. These corrected ages should be interpreted as diagnostics of the present effective torque law, not as independent source ages. This is especially important for sources such as Vela, where the present braking-index-corrected age differs substantially from the independently inferred supernova-remnant age.

\item For PSR~J0537$-$6910, the timing-based amplitude bounds converge near $\alpha\sim10^{-6}$ under the fiducial stellar model if the near-asymptotic inter-glitch behaviour is identified with an r-mode-dominated torque. In that case, the corresponding signal would lie near $83,{\rm Hz}$ with an expected strain of order $h_0\sim1$--$7\times10^{-25}$. For the fiducial stellar model used here, the corresponding strain is $h_{0,\rm fid}\approx3.8\times10^{-25}$. A continuous-wave upper limit below this level, obtained with a glitch-aware phase model, would challenge the simplest fiducial r-mode-dominated interpretation of J0537. The alternative vortex-creep interpretation remains viable in the absence of a required continuous-wave signal.

\item The idealized gravitational-wave ranking identifies Vela, the Crab, and J2229$+$6114 as the sources for which future third-generation non-detections would be most constraining under the adopted assumptions. J0537 remains the most interesting prospective detection candidate if its higher-order torque is genuinely associated with an r-mode. However, because J0537 glitches frequently, a realistic semi-coherent or glitch-limited search will be less sensitive than the idealized year-long coherent ranking.

\end{enumerate}

The main strength of this framework is that it extracts quantitative and observationally useful constraints from timing data alone. Its main limitation is that the resulting inferences remain conditional on a phenomenological four-channel model with non-negative, slowly varying effective torque coefficients. In real pulsars, magnetospheric variability, glitch recovery, superfluid coupling, and other time-dependent effects can complicate that interpretation. Even so, the framework provides a transparent first-order diagnostic, cleanly separates robust algebraic conclusions from model-dependent physical interpretation, and identifies the sources for which joint timing and gravitational-wave analysis is most promising. Future third-generation detectors, together with improved timing and thermal constraints, will provide increasingly direct tests of whether the high-order timing signatures identified here are associated with genuine r-mode gravitational-wave emission.

\section*{Acknowledgements}

The authors thank the anonymous referee for a careful reading of the
manuscript and for constructive comments that substantially improved
the presentation, uncertainty analysis, and discussion of the physical
limitations of the model.

\section*{Data availability}

The pulsar timing parameters used in this work were obtained from the ATNF Pulsar Catalogue \citep{ManchesterHobbs2005} and from the source-specific references cited in the text. All derived quantities reported in Table~\ref{tab:results} are computed from the equations given in Sections~\ref{sec:bounds} and \ref{sec:rmode_physics}, using the fiducial stellar model stated in the manuscript. The figure-generation scripts and derived tables will be made available by the corresponding author upon reasonable request.

\end{document}